\documentclass[fleqn,usenatbib]{rasti}

\usepackage{newtxtext,newtxmath}
\usepackage{algorithm}
\usepackage{algorithmic}
\usepackage[T1]{fontenc}

\DeclareRobustCommand{\VAN}[3]{#2}
\let\VANthebibliography\thebibliography
\def\thebibliography{\DeclareRobustCommand{\VAN}[3]{##3}\VANthebibliography}

\usepackage{graphicx}	
\usepackage{amsmath}	

\title[Plages Identification]{An Image Processing approach to identify solar plages observed at 393.37 nm by the Kodaikanal Solar Observatory}

\author[Sarvesh Gharat et al.]{
Sarvesh Gharat,$^{1}$\thanks{E-mail: sarveshgharat19@gmail.com}
,Bhaskar Bose$^{2}$
,Abhimanyu Borthakur$^{3}$
 and Rakesh Mazumder$^{4}$
\\
$^{1}$ Centre for Machine Intelligence and Data Science, Indian Institute of Technology Bombay, 400076, Mumbai, India\\
$^{2}$ Smart Mobility Group, Tata Consultancy Services, 560067, Bangalore, India\\
$^{3}$ Department of Electronics and Communication Engineering, Manipal Institute of Technology, 576104, Karnataka, India\\
$^{4}$  Birla Industrial \& Technological Museum, Kolkata,  National Council of Science Museums, Ministry of Culture, Govt. of India
}

\date{Accepted XXX. Received YYY; in original form ZZZ}

\pubyear{2022}

\begin{document}
\label{firstpage}
\pagerange{\pageref{firstpage}--\pageref{lastpage}}
\maketitle

\begin{abstract}
Solar plages, which are bright regions on the Sun's surface, are an important indicator of solar activity. In this study, we propose an automated algorithm for identifying solar plages in Ca K wavelength solar data obtained from the Kodaikanal Solar Observatory. The algorithm successfully annotates all visually identifiable plages in an image and outputs the corresponding calculated plage index. We perform a time series analysis of the plage index (rolling mean) across multiple solar cycles to test the algorithm's reliability and robustness. The results show a strong correlation between the calculated plage index and those reported in a previous study. The correlation coefficients obtained for all the solar cycles are higher than 0.90, indicating the reliability of the model. We also suggest that adjusting the hyperparameters appropriately for a specific image using our web-based app can increase the model's efficiency. The algorithm has been deployed on the Streamlit Community Cloud platform, where users can upload images and customize the hyperparameters for desired results. The input data used in this study is freely available from the KSO data archive, and the code and the generated data are publicly available on our GitHub repository. Our proposed algorithm provides an efficient and reliable method for identifying solar plages, which can aid the study of solar activity and its impact on the Earth's climate, technology, and space weather.
\end{abstract}

\begin{keywords}
Sun: chromosphere  -- Sun: faculae, plages -- techniques: image processing
\end{keywords}





\section{Introduction} 

The importance of space weather is becoming  more apparent in recent years \citep{2006LRSP32S}. The Sun's magnetic activity is essential in regulating space weather, but we only have continuous records of solar magnetic field observations for the last few solar cycles. To better understand the variation of solar magnetic fields, we need to reconstruct magnetic field proxies for past solar cycles. Fortunately, we have data on solar magnetic features such as sunspots, plages, and filaments for over a century, which we can use as proxies to indirectly comprehend the long-term variation of the Sun's magnetic field. Ca-K images are particularly useful for illustrating the long-term variability of the chromospheric magnetic field \citep{2018A&A...609A..92C,2021ApJ...922L..12P}. The Mount Wilson \citep{foukal2009century} and the Kodaikanal Solar Observatory \citep{2014RAA....14..229R} have century-long records of Ca-K images. Solar Plages are regions of high magnetic field concentration that can help trace the Sun's magnetic activity \citep{shine1974physical, azariadis1986sunspots, olson1978solar, neidig1989importance, mackay2008solar, canfield2000sigmoids, shine1972physical}. Although different authors may use different definitions for the Ca II K plage index, it is commonly defined as the fractional area of the total plage area to that of the full solar disk area \citep{2020ApJ...897..181B}. Ca II K plage indices show periodic variation with the solar cycle \citep{2020ApJ...897..181B,2019A&A...625A..69C,2020A&A...639A..88C,chatterjee2016butterfly}. \citep{chatterjee2016butterfly} have shown that the temporal evolution of the latitude distribution of the plages depicts a butterfly pattern, similar to the temporal evolution of the latitude distribution of the sunspots.

Precise measurements of solar irradiance are crucial for studying the climate. The Ca II K observations are highly correlated with the non-sunspot magnetic field, which can provide valuable information on the surface coverage by bright plage and network regions \citep{1955ApJ...121..349B,1975ApJ...200..747S,1985ApJ...299L..47S,2009A&A...497..273L,2016A&A...585A..40P,2016ASPC..504..227C,2017ApJS..229...12K}. As a result, historical Ca II K spectroheliograms can be used to improve long-term irradiance models significantly, especially with the availability of the digitization of various solar data archives \citep{2021A&A...656A.104C,1985LNP...233..282R,1996SoPh..167..115K,1996GeoRL..23.2169F,1998GeoRL..25.2909F,1998SoPh..177..295C,1998ApJ...496..998W,2003SoPh..214...89Z,2005MmSAI..76..862L,2009A&A...499..627E,2009ApJ...698.1000E,2009SoPh..255..239T,2010SoPh..264...31B,2011ApJ...730...51S,priyal2014long,priyal2017long,chatterjee2016butterfly,2017ApJ...841...70C}.\\
Since 1907, the Kodaikanal Solar Observatory has been observing the Sun using photographic plates with a 30 cm objective and f-ratio of f/21 in the Ca K wavelength \citep{chatterjee2016butterfly} \citep{hasan2010solar}. This has resulted in a substantial amount of data over the years, with an effective spatial resolution of around 2 arcsecs for most of the documented time. More recently, the plates have been digitized using a CCD sensor to create 4096 x 4096 raw images with 16-bit resolution \citep{2014SoPh..289..137P}.\\
Manual identification of Solar Plages for century long data is not only time consuming, it also introduces human biasing \citep{barata2018software} \citep{benkhalil2003automatic}. In this study, an image processing algorithm is proposed to identify plages from Ca II spectroheliograms obtained from the Kodaikanal Observatory. The data required for this study has been collected from the KSO data archive. Various studies have been conducted to identify different solar features such as sun spots, plages, filaments, etc \citep{barata2018software} \citep{benkhalil2003automatic} \citep{aschwanden2010image} \citep{qahwaji2005automatic} \citep{aboudarham2008automatic} \citep{scholl2008automatic}, but, there is no exclusive study conducted on the Kodaikanal Solar Observatory. Therefore, this study proposes an image processing algorithm to automate the process of plage identification specifically for the Kodaikanal Solar Observatory.\\
In their study \citep{barata2018software}, Barata et al. employed morphological transformations to segment Coimbra Observatory spectroheliograms, utilizing dilation, erosion, and the top hat operator before thresholding. Benkhalil et al. \citep{benkhalil2003automatic} used thresholding and basic morphological operations similar to \citep{barata2018software} to identify active regions in H$\alpha$ and Calcium K images from the Meudon Observatory. However, the variation in images obtained from the KSO data archive across solar cycles 16-22 made it difficult to apply such algorithms to this study. Aschwanden et al. \citep{aschwanden2010image} provided a comprehensive review of image processing for identifying different solar features and time dependency, including the use of neural networks for feature identification, but their study required a large amount of labeled data and is therefore unsuitable for this approach. Qahwaji and Colak \citep{qahwaji2005automatic} focused on identifying plages, with a proposed algorithm using morphological classification and hole filling on Meudon Observatory data to differentiate between plages and filaments. Similarly, in \citep{aboudarham2008automatic} and \citep{scholl2008automatic}, the authors used various morphological transformations to identify multiple solar features. However, due to variations in image properties such as brightness, contrast, and artificial artifacts, a novel algorithm using the OpenCV \citep{opencv_library} library is proposed for identifying plages in the Kodaikanal Solar Observatory data archive. The algorithm was tested by analyzing the time series variation of the plage index across solar cycles 16-22 and compared with the plage index reported by \citep{chatzistergos2020analysis} to confirm its efficiency.


\section{Data Description}
The Kodaikanal Solar Observatory (KSO) has been capturing photoheliograms of the Sun on a regular basis since 1904 \citep{jha2022extending} using a 15cm aperture telescope. In addition, the observatory has been capturing Ca-K line spectroheliograms since 1906 using photographic plates observed through an unaltered telescope with a 30cm objective lens and an f-ratio of f/21 \citep{chowdhury2022analysis}, resulting in an image size of about 60mm. The photographic plates were digitized using a 16-bit digitizer, allowing for high-resolution scans that preserve the details of the original images \citep{chowdhury2022analysis} \citep{chatterjee2016butterfly} \citep{chowdhury2022analysis} \citep{priyal2017long}. The spatial resolution of the obtained spectroheliograms is restricted by the prevailing seeing conditions, typically around 2 arcsecs \citep{chatterjee2016butterfly} on most days. Furthermore, the spectral window obtained from the exit slit of the spectroheliograph is 0.5 \AA \citep{chatterjee2016butterfly}, with the Ca-K line centered at 3933.67 \AA. It should be noted that there is a maximum uncertainty of 0.1 Å in the centering of the Ca-K line on the exit slit, which is due to both, the visual setting of the spectrum and the stability of the spectroheliograph \citep{priyal2014long}.\\
The KSO data archive contains Ca-K data up to October 2007 \citep{chatterjee2016butterfly}, which is categorized into three levels. The "Level 0" comprise raw images requiring preprocessing before analysis \citep{pal2020solar}, whereas "Level 1" images have undergone basic calibrations, such as bias correction, disc centering, and rotation correction with non-uniform radii \citep{pal2020solar}. Neither "Level 0" nor "Level 1" images account for the center-to-limb variation (CLV) \citep{potzi2022correction}, which is necessary to eliminate large-scale intensity variations produced by telescope optics and achieve uniform intensity and contrast. This is only addressed in the "Level 2" images \citep{pal2020solar}, which, along with disc centering and uniform radii, are the most suitable for analysis. Consequently, for this study, we limit our investigation to the "Level 2" images.

\section{Methodology}

The initial step involves loading the solar images as grayscale images using OpenCV \citep{opencv_library}, after which the algorithm proceeds with the following steps:

\subsection{CLAHE}
To enhance the contrast of the images, a technique called Contrast Limited Adaptive Histogram Equalization (CLAHE) was utilized \citep{reza2004realization}. This method is an extension of the commonly used histogram equalization technique \citep{pizer1987adaptive}, which has been shown to improve image contrast \citep{abdullah2007dynamic}. However, global histogram equalization techniques are not suitable for our study due to inconsistent pixel intensities and varying brightness levels in the images. Therefore, CLAHE was chosen as it is based on the local intensity statistics. To prevent over-amplification of noise, we used a clipping factor \citep{sundaram2011histogram} \citep{liu2019adaptive} as one of the parameters that require experimental tuning to achieve the best results. After experimentation, a clipping factor of 2 was found to produce the most satisfactory outcomes. \\

\subsection{Image Thresholding}


Image thresholding plays a crucial role in detecting plages by segmenting the image based on a predetermined threshold value \citep{chowdhury1995image}. In our research, a constant threshold value of 180 was employed for the majority of cases, which proved to be effective due to the utilization of CLAHE \citep{reza2004realization}. However, it is worth noting that in some instances, a threshold value as low as 100, or as high as 220 may yield superior results. This conclusion was reached after careful analysis of all the archived images. As a result, our web application provides users with a variable threshold slider that enables them to select the most appropriate threshold value for a specific image.


\subsection{Opening}

Opening \citep{chen1995recursive} is a morphological image processing technique that involves the use of a structuring element \citep{song1990analysis} to eliminate small objects or noise from an input image. The structuring element is a small binary image that determines the shape and size of the objects to be removed. The opening operation works by sliding the structuring element over the input image and performing a logical AND operation between the structuring element and the corresponding pixels in the input image \citep{chen1995recursive}. If all the pixels in the structuring element are white (i.e., have a value of 1) and the corresponding pixels in the input image are white, then the output pixel value is set to 1. If not, the output pixel value is black (i.e., has a value of 0). In our study, we used a 3x3 elliptical kernel \citep{landstrom2013adaptive, tian2016optimization} as the structuring element, which was chosen after experimenting with several other kernels.

\subsection{Erosion and Dilation}
Erosion and dilation are basic image processing transformations commonly used to enhance and remove small contrasting spots in images \citep{soille2004erosion}. The erosion operation removes regions that are smaller than the size of the structuring element, while dilation joins two regions where the distance between them is less than or equal to the size of the structuring element. The structuring element used in this step is identical to that of section 3.3 - a 3x3 elliptical kernel. In our algorithm, erosion and dilation are important for reducing the noise in the images. In this study, we apply two iterations of erosion and one iteration of dilation, after the opening operation is performed on the thresholded image.

\subsection{Area Filtering}
the image obtained in section 3.4 using pre-built functions available in OpenCV \citep{opencv_library} and determining the area of the solar disc. With the area of the solar disc, the radius, center, heliographic latitude, longitude, polar angle, and semi-diameter are calculated according to the methodology outlined in \citep{hiremath2019kodaikanal}. Using the calculated semi-diameter and solar disc radius, the pixel size and area are computed using the formula given in 
\citep{pucha2016development}\\
Next, the contours are extracted from the segmented image, and the area corresponding to each contour (which either corresponds to a plage or noise incorrectly identified as a plage), is calculated using the steps described in \citep{hiremath2019kodaikanal}. Contours that are too small or too large are very likely to be noisy components, and are hence, discarded using a lower and an upper threshold, respectively. Only contours with an area in between the two thresholds are retained in the output image. Although this method of hard thresholding can lead to the elimination of some plages, the thresholds have been determined through experimentation on multiple images. Moreover, users can also set their own threshold values based on their prior knowledge of the solar cycle to avoid losing any plages.\\ \\
In summary, the proposed algorithm is comprised of several steps. Firstly, the input image is read in grayscale and enhanced with Contrast Limited Adaptive Histogram Equalization (CLAHE) to improve contrast. The algorithm allows the user to customize the clipping factor for CLAHE. Then, the image is thresholded using a user-defined threshold value. An elliptical structuring element of size 3x3 is initialized and used for opening, followed by two iterations of erosion and one iteration of dilation. The resulting image is used to extract contours using cv2.findContours, and the area of each contour is calculated while accounting for the projectional effects of the sun's spherical shape. The algorithm retains only the contours with an area within the user-defined lower and upper area thresholds and discards others. These retained contours are used to calculate the plage index and are overlaid on the input image to highlight the plages using cv2.drawContours. Finally, the output of the algorithm is an image highlighting the plages and the calculated plage index. In the event that the user fails to provide appropriate hyperparameters, the algorithm defaults to the parameters listed in Table \ref{table}.
\begin{table}
    \caption{Default Values of Hyperparameters}
    \label {table}
    \begin{tabular}{ ||c | c || }
    \hline
    \hline
    Hyperparameter & Default Value \\ [0.5 ex]
    \hline \hline
    CLAHE & 2  \\
    Binary Threshold & 180 \\
    Lower Area Limit & 0 \\ 
    Upper Area Limit & 1000\\ 
    \hline
    \hline
    \end{tabular}
    \centering
\end{table}



\section{Results and Discussion}
We used the automated algorithm for identifying solar plages on the Ca K wavelength solar data obtained from the Kodaikanal Solar Observatory, for multiple solar cycles ranging from Cycle 16 to Cycle 22, to confirm the validity of our approach. The number of images utilized for testing the algorithm in each cycle is listed in Table \ref{table:1}.
\begin{table}
    \caption{Number of images for each solar cycle from the KSO archive}
    \label {table:1}
    \begin{tabular}{ ||c | c || }
    \hline
    \hline
    Solar Cycle & Number of Images \\ [0.5 ex]
    \hline \hline
    Solar Cycle 16 & 989  \\
    Solar Cycle 17 & 969 \\
    Solar Cycle 18 & 1000 \\
    Solar Cycle 19 & 999 \\
    Solar Cycle 20 & 1000 \\
    Solar Cycle 21 & 1052 \\
    Solar Cycle 22 & 939 \\
    \hline
    Total & 6948 \\
    \hline
    \hline
    \end{tabular}
    \centering
\end{table}

The proposed algorithm can only process images from the Kodaikanal Observatory, which should be in .jpg format. To make it more convenient for users, the algorithm has been deployed on the Streamlit Community Cloud platform (https://abhimanyu911-plages-identification-app-yfyqbq.streamlit.app/), where users can upload images and customize the hyperparameters for desired results. The users can experiment with various parameters such as clip limit (ranging from 1 to 10), lower area threshold (ranging from 0 to 500), upper area threshold (ranging from 0 to 10000), and binary threshold (ranging from 0 to 255) and see the corresponding effect on the input image in real-time. This feature allows users to visualize and understand how different hyperparameters impact the output. The algorithm not only overlays the plage contour on the input image but also provides the calculated plage index, which can be compared with the results obtained in \citep{chatzistergos2020analysis}(if available).

\begin{figure}
    \centering
    \includegraphics[scale = 0.3]{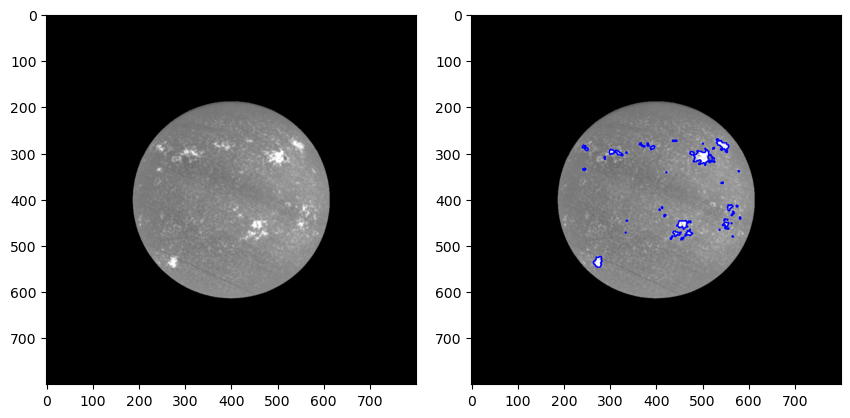}
    \caption{Automatic identification of Solar Plages using the proposed algorithm}
    \label{fig:fig1}
\end{figure}

The input image in Figure \ref{fig:fig1} is an example of the proposed algorithm's performance. The algorithm successfully annotates all visually identifiable plages in the image and also outputs the corresponding calculated plage index of $0.02733$. To test the algorithm's reliability and robustness, we perform a time series analysis of the plage index (rolling mean) across multiple solar cycles. Additionally, we calculate the correlation coefficient, which measures the strength of a linear relationship between two variables, and the $R^2$ score, which indicates the proportion of the variation in the dependent variable (calculated plage index) that can be predicted from the independent variable (plage index from \citep{chatzistergos2020analysis}).

\begin{figure}
    \centering
    \includegraphics[scale = 0.4]{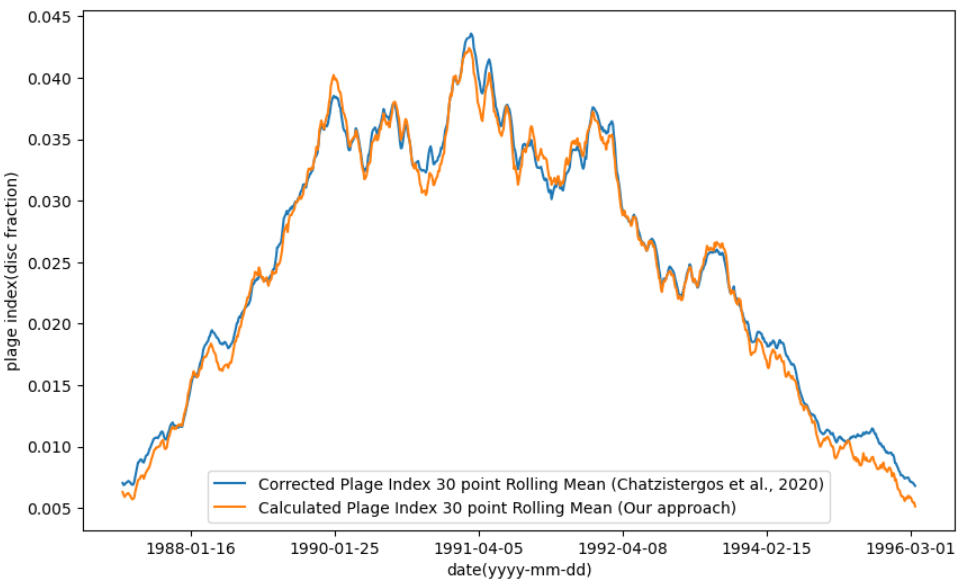}
    \caption{30 points rolling mean of plage index in Solar Cycle 22}
    \label{fig:fig2}
\end{figure}

The results obtained for the Solar Cycle 22 are presented in Figure \ref{fig:fig2}, which shows a strong correlation between the calculated plage index and that reported in \citep{chatzistergos2020analysis}. This correlation is confirmed by the high values of the correlation coefficient between the two variables. Table \ref{table:2} provides a summary of the correlation coefficients and $R^2$ values obtained for different solar cycles.

\begin{table}
    \caption{Correlation Coefficient and $R^2$ values across multiple solar cycles}
    \label {table:2}
    \begin{tabular}{ ||c | c | c || }
    \hline
    \hline
    Solar Cycle & Correlation Coefficient & $R^2$ \\ [0.5 ex]
    \hline \hline
    Solar Cycle 16 & 0.91 & 0.83  \\
    Solar Cycle 17 & 0.95 & 0.91 \\
    Solar Cycle 18 & 0.95 & 0.91 \\
    Solar Cycle 19 & 0.94 & 0.88\\
    Solar Cycle 20 & 0.96 & 0.92 \\
    Solar Cycle 21 & 0.97 & 0.95 \\
    Solar Cycle 22 & 0.94 & 0.89 \\
    \hline
    \hline
    \end{tabular}
    \centering
\end{table}

Table \ref{table:2} demonstrates that the correlation coefficient for all the solar cycles is higher than 0.90, indicating the reliability of the model. We also suggest that adjusting the hyperparameters appropriately for each image using our web-based app can increase the model's efficiency and result in better outputs. However, this process may be time-consuming in case a large number of images are present for annotation.

\section{Conclusion}
In conclusion, we present an automated algorithm for identifying solar plages in Ca K wavelength solar data obtained from the Kodaikanal Solar Observatory. The proposed algorithm is able to successfully annotate visually identifiable plages and calculate the corresponding plage index, which indicates the level of magnetic activity present in the Sun. To test the reliability and robustness of our algorithm, we performed a time series analysis of the calculated plage index (rolling mean) across multiple solar cycles. The results showed a high correlation between the calculated values and those reported in \citep{chatzistergos2020analysis}, as indicated by the high values of the correlation coefficient and $R^2$ scores.\\
Furthermore, to make the algorithm more accessible to users, we have deployed it on the Streamlit Community Cloud platform, where users can upload images and customize the hyperparameters for desired results. This feature allows users to experiment with different hyperparameters and visualize their impact on the output image in real time. We believe that our proposed algorithm has the potential to improve the accuracy and efficiency of identifying solar plages, which is essential for studying the Sun's magnetic activity and its effects on space weather. Furthermore, the availability of the input data and the code used in this study at public repositories ensures the reproducibility and transparency of our results.

\section*{Acknowledgements}
We want to thank the Kodaikanal solar observatory, a facility of the Indian Institute of Astrophysics, Bangalore, India for making the archival data available. This data is available for public use at http://kso.iiap.res.in through a service developed at IUCAA under the Data Driven Initiatives project funded by the National Knowledge Network\\
The authors would like to express their gratitude to the anonymous referees for their valuable suggestions, which have significantly improved the quality of this study.
\section*{Data Availability}

The data used in this study can be obtained for free from the Kodaikanal Solar Observatory (KSO) data archive website (https://kso.iiap.res.in/data). In addition, we have made our code and the resulting data publicly accessible on our GitHub repository (https://github.com/SarveshVGharat/Plages-Identification).



\bibliographystyle{mnras}
\bibliography{example} 

\begin{thebibliography}{}
\makeatletter
\relax
\def\mn@urlcharsother{\let\do\@makeother \do\$\do\&\do\#\do\^\do\_\do\%\do\~}
\def\mn@doi{\begingroup\mn@urlcharsother \@ifnextchar [ {\mn@doi@}
  {\mn@doi@[]}}
\def\mn@doi@[#1]#2{\def\@tempa{#1}\ifx\@tempa\@empty \href
  {http://dx.doi.org/#2} {doi:#2}\else \href {http://dx.doi.org/#2} {#1}\fi
  \endgroup}
\def\mn@eprint#1#2{\mn@eprint@#1:#2::\@nil}
\def\mn@eprint@arXiv#1{\href {http://arxiv.org/abs/#1} {{\tt arXiv:#1}}}
\def\mn@eprint@dblp#1{\href {http://dblp.uni-trier.de/rec/bibtex/#1.xml}
  {dblp:#1}}
\def\mn@eprint@#1:#2:#3:#4\@nil{\def\@tempa {#1}\def\@tempb {#2}\def\@tempc
  {#3}\ifx \@tempc \@empty \let \@tempc \@tempb \let \@tempb \@tempa \fi \ifx
  \@tempb \@empty \def\@tempb {arXiv}\fi \@ifundefined
  {mn@eprint@\@tempb}{\@tempb:\@tempc}{\expandafter \expandafter \csname
  mn@eprint@\@tempb\endcsname \expandafter{\@tempc}}}

\bibitem[\protect\citeauthoryear{Abdullah-Al-Wadud, Kabir, Dewan  \&
  Chae}{Abdullah-Al-Wadud et~al.}{2007}]{abdullah2007dynamic}
Abdullah-Al-Wadud M.,  Kabir M.~H.,  Dewan M. A.~A.,   Chae O.,  2007, IEEE
  Transactions on Consumer Electronics, 53, 593

\bibitem[\protect\citeauthoryear{Aboudarham, Scholl, Fuller, Fouesneau,
  Galametz, Gonon, Maire  \& Leroy}{Aboudarham
  et~al.}{2008}]{aboudarham2008automatic}
Aboudarham J.,  Scholl I.,  Fuller N.,  Fouesneau M.,  Galametz M.,  Gonon F.,
  Maire A.,   Leroy Y.,  2008, in Annales Geophysicae. pp 243--248

\bibitem[\protect\citeauthoryear{Aschwanden}{Aschwanden}{2010}]{aschwanden2010image}
Aschwanden M.~J.,  2010, Solar Physics, 262, 235

\bibitem[\protect\citeauthoryear{Azariadis \& Guesnerie}{Azariadis \&
  Guesnerie}{1986}]{azariadis1986sunspots}
Azariadis C.,  Guesnerie R.,  1986, The review of economic studies, 53, 725

\bibitem[\protect\citeauthoryear{{Babcock} \& {Babcock}}{{Babcock} \&
  {Babcock}}{1955}]{1955ApJ...121..349B}
{Babcock} H.~W.,  {Babcock} H.~D.,  1955, \mn@doi [\apj] {10.1086/145994},
  \href {https://ui.adsabs.harvard.edu/abs/1955ApJ...121..349B} {121, 349}

\bibitem[\protect\citeauthoryear{Barata, Carvalho, Dorotovi{\v{c}}, Pinheiro,
  Garcia, Fernandes  \& Louren{\c{c}}o}{Barata
  et~al.}{2018}]{barata2018software}
Barata T.,  Carvalho S.,  Dorotovi{\v{c}} I.,  Pinheiro F.~J.,  Garcia A.,
  Fernandes J.,   Louren{\c{c}}o A.~M.,  2018, Astronomy and Computing, 24, 70

\bibitem[\protect\citeauthoryear{Benkhalil, Zharkova, Ipson  \&
  Zharkov}{Benkhalil et~al.}{2003}]{benkhalil2003automatic}
Benkhalil A.,  Zharkova V.,  Ipson S.,   Zharkov S.,  2003, in Proceedings of
  the AISB. pp 66--73

\bibitem[\protect\citeauthoryear{{Bertello}, {Ulrich}  \& {Boyden}}{{Bertello}
  et~al.}{2010}]{2010SoPh..264...31B}
{Bertello} L.,  {Ulrich} R.~K.,   {Boyden} J.~E.,  2010, \mn@doi [\solphys]
  {10.1007/s11207-010-9570-z}, \href
  {https://ui.adsabs.harvard.edu/abs/2010SoPh..264...31B} {264, 31}

\bibitem[\protect\citeauthoryear{{Bertello}, {Pevtsov}  \& {Ulrich}}{{Bertello}
  et~al.}{2020}]{2020ApJ...897..181B}
{Bertello} L.,  {Pevtsov} A.~A.,   {Ulrich} R.~K.,  2020, \mn@doi [\apj]
  {10.3847/1538-4357/ab9746}, \href
  {https://ui.adsabs.harvard.edu/abs/2020ApJ...897..181B} {897, 181}

\bibitem[\protect\citeauthoryear{Bradski}{Bradski}{2000}]{opencv_library}
Bradski G.,  2000, Dr. Dobb's Journal of Software Tools

\bibitem[\protect\citeauthoryear{{Caccin}, {Ermolli}, {Fofi}  \&
  {Sambuco}}{{Caccin} et~al.}{1998}]{1998SoPh..177..295C}
{Caccin} B.,  {Ermolli} I.,  {Fofi} M.,   {Sambuco} A.~M.,  1998, \mn@doi
  [\solphys] {10.1023/A:1004938412420}, \href
  {https://ui.adsabs.harvard.edu/abs/1998SoPh..177..295C} {177, 295}

\bibitem[\protect\citeauthoryear{Canfield, Hudson  \& Pevtsov}{Canfield
  et~al.}{2000}]{canfield2000sigmoids}
Canfield R.~C.,  Hudson H.~S.,   Pevtsov A.~A.,  2000, IEEE transactions on
  plasma science, 28, 1786

\bibitem[\protect\citeauthoryear{Chatterjee, Banerjee  \& Ravindra}{Chatterjee
  et~al.}{2016}]{chatterjee2016butterfly}
Chatterjee S.,  Banerjee D.,   Ravindra B.,  2016, The Astrophysical Journal,
  827, 87

\bibitem[\protect\citeauthoryear{{Chatterjee}, {Mandal}  \&
  {Banerjee}}{{Chatterjee} et~al.}{2017}]{2017ApJ...841...70C}
{Chatterjee} S.,  {Mandal} S.,   {Banerjee} D.,  2017, \mn@doi [\apj]
  {10.3847/1538-4357/aa709d}, \href
  {https://ui.adsabs.harvard.edu/abs/2017ApJ...841...70C} {841, 70}

\bibitem[\protect\citeauthoryear{{Chatzistergos}, {Ermolli}, {Solanki}  \&
  {Krivova}}{{Chatzistergos} et~al.}{2016}]{2016ASPC..504..227C}
{Chatzistergos} T.,  {Ermolli} I.,  {Solanki} S.~K.,   {Krivova} N.~A.,  2016,
  in {Dorotovic} I.,  {Fischer} C.~E.,   {Temmer} M.,  eds,  Astronomical
  Society of the Pacific Conference Series Vol. 504, Coimbra Solar Physics
  Meeting: Ground-based Solar Observations in the Space Instrumentation Era.
  p.~227

\bibitem[\protect\citeauthoryear{{Chatzistergos}, {Ermolli}, {Solanki}  \&
  {Krivova}}{{Chatzistergos} et~al.}{2018}]{2018A&A...609A..92C}
{Chatzistergos} T.,  {Ermolli} I.,  {Solanki} S.~K.,   {Krivova} N.~A.,  2018,
  \mn@doi [\aap] {10.1051/0004-6361/201731511}, \href
  {https://ui.adsabs.harvard.edu/abs/2018A&A...609A..92C} {609, A92}

\bibitem[\protect\citeauthoryear{{Chatzistergos}, {Ermolli}, {Krivova}  \&
  {Solanki}}{{Chatzistergos} et~al.}{2019}]{2019A&A...625A..69C}
{Chatzistergos} T.,  {Ermolli} I.,  {Krivova} N.~A.,   {Solanki} S.~K.,  2019,
  \mn@doi [\aap] {10.1051/0004-6361/201834402}, \href
  {https://ui.adsabs.harvard.edu/abs/2019A&A...625A..69C} {625, A69}

\bibitem[\protect\citeauthoryear{Chatzistergos et~al.,}{Chatzistergos
  et~al.}{2020b}]{chatzistergos2020analysis}
Chatzistergos T.,  et~al., 2020b, Astronomy \& Astrophysics, 639, A88

\bibitem[\protect\citeauthoryear{{Chatzistergos} et~al.,}{{Chatzistergos}
  et~al.}{2020a}]{2020A&A...639A..88C}
{Chatzistergos} T.,  et~al., 2020a, \mn@doi [\aap]
  {10.1051/0004-6361/202037746}, \href
  {https://ui.adsabs.harvard.edu/abs/2020A&A...639A..88C} {639, A88}

\bibitem[\protect\citeauthoryear{{Chatzistergos}, {Krivova}, {Ermolli}, {Yeo},
  {Mandal}, {Solanki}, {Kopp}  \& {Malherbe}}{{Chatzistergos}
  et~al.}{2021}]{2021A&A...656A.104C}
{Chatzistergos} T.,  {Krivova} N.~A.,  {Ermolli} I.,  {Yeo} K.~L.,  {Mandal}
  S.,  {Solanki} S.~K.,  {Kopp} G.,   {Malherbe} J.-M.,  2021, \mn@doi [\aap]
  {10.1051/0004-6361/202141516}, \href
  {https://ui.adsabs.harvard.edu/abs/2021A&A...656A.104C} {656, A104}

\bibitem[\protect\citeauthoryear{Chen \& Haralick}{Chen \&
  Haralick}{1995}]{chen1995recursive}
Chen S.,  Haralick R.~M.,  1995, IEEE Transactions on image processing, 4, 335

\bibitem[\protect\citeauthoryear{Chowdhury \& Little}{Chowdhury \&
  Little}{1995}]{chowdhury1995image}
Chowdhury M.~H.,  Little W.~D.,  1995, in IEEE pacific Rim conference on
  communications, computers, and signal processing. Proceedings. pp 585--589

\bibitem[\protect\citeauthoryear{Chowdhury, Belur, Bertello  \&
  Pevtsov}{Chowdhury et~al.}{2022}]{chowdhury2022analysis}
Chowdhury P.,  Belur R.,  Bertello L.,   Pevtsov A.~A.,  2022, The
  Astrophysical Journal, 925, 81

\bibitem[\protect\citeauthoryear{{Ermolli}, {Marchei}, {Centrone}, {Criscuoli},
  {Giorgi}  \& {Perna}}{{Ermolli} et~al.}{2009a}]{2009A&A...499..627E}
{Ermolli} I.,  {Marchei} E.,  {Centrone} M.,  {Criscuoli} S.,  {Giorgi} F.,
  {Perna} C.,  2009a, \mn@doi [\aap] {10.1051/0004-6361/200811406}, \href
  {https://ui.adsabs.harvard.edu/abs/2009A&A...499..627E} {499, 627}

\bibitem[\protect\citeauthoryear{{Ermolli}, {Solanki}, {Tlatov}, {Krivova},
  {Ulrich}  \& {Singh}}{{Ermolli} et~al.}{2009b}]{2009ApJ...698.1000E}
{Ermolli} I.,  {Solanki} S.~K.,  {Tlatov} A.~G.,  {Krivova} N.~A.,  {Ulrich}
  R.~K.,   {Singh} J.,  2009b, \mn@doi [\apj] {10.1088/0004-637X/698/2/1000},
  \href {https://ui.adsabs.harvard.edu/abs/2009ApJ...698.1000E} {698, 1000}

\bibitem[\protect\citeauthoryear{{Foukal}}{{Foukal}}{1996}]{1996GeoRL..23.2169F}
{Foukal} P.,  1996, \mn@doi [\grl] {10.1029/96GL01356}, \href
  {https://ui.adsabs.harvard.edu/abs/1996GeoRL..23.2169F} {23, 2169}

\bibitem[\protect\citeauthoryear{{Foukal}}{{Foukal}}{1998}]{1998GeoRL..25.2909F}
{Foukal} P.,  1998, \mn@doi [\grl] {10.1029/98GL02057}, \href
  {https://ui.adsabs.harvard.edu/abs/1998GeoRL..25.2909F} {25, 2909}

\bibitem[\protect\citeauthoryear{Foukal, Bertello, Livingston, Pevtsov, Singh,
  Tlatov  \& Ulrich}{Foukal et~al.}{2009}]{foukal2009century}
Foukal P.,  Bertello L.,  Livingston W.~C.,  Pevtsov A.~A.,  Singh J.,  Tlatov
  A.~G.,   Ulrich R.~K.,  2009, Solar Physics, 255, 229

\bibitem[\protect\citeauthoryear{Hasan, Mallik, Bagare  \& Rajaguru}{Hasan
  et~al.}{2010}]{hasan2010solar}
Hasan S.,  Mallik D.,  Bagare S.,   Rajaguru S.,  2010, in , Magnetic Coupling
  between the Interior and Atmosphere of the Sun.
Springer, pp 12--36

\bibitem[\protect\citeauthoryear{Hiremath, Krishna, Chinmaya, Gurumath
  et~al.}{Hiremath et~al.}{2019}]{hiremath2019kodaikanal}
Hiremath K.,  Krishna S.,  Chinmaya S.,  Gurumath S.~R.,   et~al., 2019, arXiv
  preprint arXiv:1909.00406

\bibitem[\protect\citeauthoryear{Jha, Hegde, Priyadarshi, Mandal, Ravindra  \&
  Banerjee}{Jha et~al.}{2022}]{jha2022extending}
Jha B.~K.,  Hegde M.,  Priyadarshi A.,  Mandal S.,  Ravindra B.,   Banerjee D.,
   2022, arXiv preprint arXiv:2210.06922

\bibitem[\protect\citeauthoryear{{Kahil}, {Riethm{\"u}ller}  \&
  {Solanki}}{{Kahil} et~al.}{2017}]{2017ApJS..229...12K}
{Kahil} F.,  {Riethm{\"u}ller} T.~L.,   {Solanki} S.~K.,  2017, \mn@doi [\apjs]
  {10.3847/1538-4365/229/1/12}, \href
  {https://ui.adsabs.harvard.edu/abs/2017ApJS..229...12K} {229, 12}

\bibitem[\protect\citeauthoryear{{Kariyappa} \& {Pap}}{{Kariyappa} \&
  {Pap}}{1996}]{1996SoPh..167..115K}
{Kariyappa} R.,  {Pap} J.~M.,  1996, \mn@doi [\solphys] {10.1007/BF00146331},
  \href {https://ui.adsabs.harvard.edu/abs/1996SoPh..167..115K} {167, 115}

\bibitem[\protect\citeauthoryear{Landstr{\"o}m \& Thurley}{Landstr{\"o}m \&
  Thurley}{2013}]{landstrom2013adaptive}
Landstr{\"o}m A.,  Thurley M.~J.,  2013, Pattern Recognition Letters, 34, 1416

\bibitem[\protect\citeauthoryear{{Lefebvre} et~al.,}{{Lefebvre}
  et~al.}{2005}]{2005MmSAI..76..862L}
{Lefebvre} S.,  et~al., 2005, \memsai, \href
  {https://ui.adsabs.harvard.edu/abs/2005MmSAI..76..862L} {76, 862}

\bibitem[\protect\citeauthoryear{Liu, Sui, Liu, Kuang, Gu  \& Chen}{Liu
  et~al.}{2019}]{liu2019adaptive}
Liu C.,  Sui X.,  Liu Y.,  Kuang X.,  Gu G.,   Chen Q.,  2019, Journal of
  Modern Optics, 66, 1590

\bibitem[\protect\citeauthoryear{{Loukitcheva}, {Solanki}  \&
  {White}}{{Loukitcheva} et~al.}{2009}]{2009A&A...497..273L}
{Loukitcheva} M.,  {Solanki} S.~K.,   {White} S.~M.,  2009, \mn@doi [\aap]
  {10.1051/0004-6361/200811133}, \href
  {https://ui.adsabs.harvard.edu/abs/2009A&A...497..273L} {497, 273}

\bibitem[\protect\citeauthoryear{Mackay, Gaizauskas  \& Yeates}{Mackay
  et~al.}{2008}]{mackay2008solar}
Mackay D.~H.,  Gaizauskas V.,   Yeates A.~R.,  2008, Solar Physics, 248, 51

\bibitem[\protect\citeauthoryear{Neidig}{Neidig}{1989}]{neidig1989importance}
Neidig D.~F.,  1989, in , Solar and Stellar Flares.
Springer, pp 261--269

\bibitem[\protect\citeauthoryear{OLSON, ROBERTS, PRINCE  \& HEDEMAN}{OLSON
  et~al.}{1978}]{olson1978solar}
OLSON R.~H.,  ROBERTS W.~O.,  PRINCE H.~D.,   HEDEMAN E.,  1978, Nature, 274,
  140

\bibitem[\protect\citeauthoryear{Pal, Verma, Rendtel, Manrique, Enke  \&
  Denker}{Pal et~al.}{2020}]{pal2020solar}
Pal P.~S.,  Verma M.,  Rendtel J.,  Manrique S. J.~G.,  Enke H.,   Denker C.,
  2020, Astronomische Nachrichten, 341, 575

\bibitem[\protect\citeauthoryear{{Penza}, {Berrilli}, {Bertello}, {Cantoresi}
  \& {Criscuoli}}{{Penza} et~al.}{2021}]{2021ApJ...922L..12P}
{Penza} V.,  {Berrilli} F.,  {Bertello} L.,  {Cantoresi} M.,   {Criscuoli} S.,
  2021, \mn@doi [\apjl] {10.3847/2041-8213/ac3663}, \href
  {https://ui.adsabs.harvard.edu/abs/2021ApJ...922L..12P} {922, L12}

\bibitem[\protect\citeauthoryear{{Pevtsov}, {Virtanen}, {Mursula}, {Tlatov}  \&
  {Bertello}}{{Pevtsov} et~al.}{2016}]{2016A&A...585A..40P}
{Pevtsov} A.~A.,  {Virtanen} I.,  {Mursula} K.,  {Tlatov} A.,   {Bertello} L.,
  2016, \mn@doi [\aap] {10.1051/0004-6361/201526620}, \href
  {https://ui.adsabs.harvard.edu/abs/2016A&A...585A..40P} {585, A40}

\bibitem[\protect\citeauthoryear{Pizer et~al.,}{Pizer
  et~al.}{1987}]{pizer1987adaptive}
Pizer S.~M.,  et~al., 1987, Computer vision, graphics, and image processing,
  39, 355

\bibitem[\protect\citeauthoryear{P{\"o}tzi, Veronig, Jarolim,
  Rodr{\'\i}guez~G{\'o}mez, Podladchikova, Baumgartner, Freislich  \&
  Strutzmann}{P{\"o}tzi et~al.}{2022}]{potzi2022correction}
P{\"o}tzi W.,  Veronig A.,  Jarolim R.,  Rodr{\'\i}guez~G{\'o}mez J.~M.,
  Podladchikova T.,  Baumgartner D.,  Freislich H.,   Strutzmann H.,  2022,
  Solar Physics, 297, 1

\bibitem[\protect\citeauthoryear{Priyal, Singh, Ravindra, Priya  \&
  Amareswari}{Priyal et~al.}{2014a}]{priyal2014long}
Priyal M.,  Singh J.,  Ravindra B.,  Priya T.,   Amareswari K.,  2014a, Solar
  Physics, 289, 137

\bibitem[\protect\citeauthoryear{{Priyal}, {Singh}, {Ravindra}, {Priya}  \&
  {Amareswari}}{{Priyal} et~al.}{2014b}]{2014SoPh..289..137P}
{Priyal} M.,  {Singh} J.,  {Ravindra} B.,  {Priya} T.~G.,   {Amareswari} K.,
  2014b, \mn@doi [\solphys] {10.1007/s11207-013-0315-7}, \href
  {https://ui.adsabs.harvard.edu/abs/2014SoPh..289..137P} {289, 137}

\bibitem[\protect\citeauthoryear{Priyal, Singh, Belur  \& Rathina}{Priyal
  et~al.}{2017}]{priyal2017long}
Priyal M.,  Singh J.,  Belur R.,   Rathina S.~K.,  2017, Solar Physics, 292, 1

\bibitem[\protect\citeauthoryear{Pucha, Hiremath  \& Gurumath}{Pucha
  et~al.}{2016}]{pucha2016development}
Pucha R.,  Hiremath K.,   Gurumath S.~R.,  2016, Journal of Astrophysics and
  Astronomy, 37, 1

\bibitem[\protect\citeauthoryear{Qahwaji \& Colak}{Qahwaji \&
  Colak}{2005}]{qahwaji2005automatic}
Qahwaji R.,  Colak T.,  2005, International Journal of Imaging Systems and
  Technology, 15, 199

\bibitem[\protect\citeauthoryear{{Raju} \& {Singh}}{{Raju} \&
  {Singh}}{2014}]{2014RAA....14..229R}
{Raju} K.~P.,  {Singh} J.,  2014, \mn@doi [Research in Astronomy and
  Astrophysics] {10.1088/1674-4527/14/2/010}, \href
  {https://ui.adsabs.harvard.edu/abs/2014RAA....14..229R} {14, 229}

\bibitem[\protect\citeauthoryear{Reza}{Reza}{2004}]{reza2004realization}
Reza A.~M.,  2004, Journal of VLSI signal processing systems for signal, image
  and video technology, 38, 35

\bibitem[\protect\citeauthoryear{{Ribes} \& {Mein}}{{Ribes} \&
  {Mein}}{1985}]{1985LNP...233..282R}
{Ribes} E.,  {Mein} P.,  1985, in {Muller} R.,  ed., , Vol.~223, High
  Resolution in Solar Physics.
p.~282, \mn@doi{10.1007/BFb0022425}

\bibitem[\protect\citeauthoryear{{Saar} \& {Linsky}}{{Saar} \&
  {Linsky}}{1985}]{1985ApJ...299L..47S}
{Saar} S.~H.,  {Linsky} J.~L.,  1985, \mn@doi [\apjl] {10.1086/184578}, \href
  {https://ui.adsabs.harvard.edu/abs/1985ApJ...299L..47S} {299, L47}

\bibitem[\protect\citeauthoryear{Scholl \& Habbal}{Scholl \&
  Habbal}{2008}]{scholl2008automatic}
Scholl I.~F.,  Habbal S.~R.,  2008, Solar Physics, 248, 425

\bibitem[\protect\citeauthoryear{{Schwenn}}{{Schwenn}}{2006}]{2006LRSP32S}
{Schwenn} R.,  2006, \mn@doi [Living Reviews in Solar Physics]
  {10.12942/lrsp-2006-2}, \href
  {https://ui.adsabs.harvard.edu/abs/2006LRSP....3....2S} {3, 2}

\bibitem[\protect\citeauthoryear{{Sheeley}, {Cooper}  \& {Anderson}}{{Sheeley}
  et~al.}{2011}]{2011ApJ...730...51S}
{Sheeley} N.~R. J.,  {Cooper} T.~J.,   {Anderson} J.~R.~L.,  2011, \mn@doi
  [\apj] {10.1088/0004-637X/730/1/51}, \href
  {https://ui.adsabs.harvard.edu/abs/2011ApJ...730...51S} {730, 51}

\bibitem[\protect\citeauthoryear{Shine \& Linsky}{Shine \&
  Linsky}{1972}]{shine1972physical}
Shine R.~A.,  Linsky J.~L.,  1972, Solar Physics, 25, 357

\bibitem[\protect\citeauthoryear{Shine \& Linsky}{Shine \&
  Linsky}{1974}]{shine1974physical}
Shine R.~A.,  Linsky J.~L.,  1974, Solar Physics, 39, 49

\bibitem[\protect\citeauthoryear{{Skumanich}, {Smythe}  \&
  {Frazier}}{{Skumanich} et~al.}{1975}]{1975ApJ...200..747S}
{Skumanich} A.,  {Smythe} C.,   {Frazier} E.~N.,  1975, \mn@doi [\apj]
  {10.1086/153846}, \href
  {https://ui.adsabs.harvard.edu/abs/1975ApJ...200..747S} {200, 747}

\bibitem[\protect\citeauthoryear{Soille}{Soille}{2004}]{soille2004erosion}
Soille P.,  2004, in , Morphological Image Analysis.
Springer, pp 63--103

\bibitem[\protect\citeauthoryear{Song \& Delp}{Song \&
  Delp}{1990}]{song1990analysis}
Song J.,  Delp E.~J.,  1990, Computer Vision, Graphics, and Image Processing,
  50, 308

\bibitem[\protect\citeauthoryear{Sundaram, Ramar, Arumugam  \& Prabin}{Sundaram
  et~al.}{2011}]{sundaram2011histogram}
Sundaram M.,  Ramar K.,  Arumugam N.,   Prabin G.,  2011, in 2011 International
  conference on signal processing, communication, computing and networking
  technologies. pp 842--846

\bibitem[\protect\citeauthoryear{Tian \& Yang}{Tian \&
  Yang}{2016}]{tian2016optimization}
Tian M.-L.,  Yang J.-M.,  2016, International Journal of Pattern Recognition
  and Artificial Intelligence, 30, 1654002

\bibitem[\protect\citeauthoryear{{Tlatov}, {Pevtsov}  \& {Singh}}{{Tlatov}
  et~al.}{2009}]{2009SoPh..255..239T}
{Tlatov} A.~G.,  {Pevtsov} A.~A.,   {Singh} J.,  2009, \mn@doi [\solphys]
  {10.1007/s11207-009-9326-9}, \href
  {https://ui.adsabs.harvard.edu/abs/2009SoPh..255..239T} {255, 239}

\bibitem[\protect\citeauthoryear{{Worden}, {White}  \& {Woods}}{{Worden}
  et~al.}{1998}]{1998ApJ...496..998W}
{Worden} J.~R.,  {White} O.~R.,   {Woods} T.~N.,  1998, \mn@doi [\apj]
  {10.1086/305392}, \href
  {https://ui.adsabs.harvard.edu/abs/1998ApJ...496..998W} {496, 998}

\bibitem[\protect\citeauthoryear{{Zharkova}, {Ipson}, {Zharkov}, {Benkhalil},
  {Aboudarham}  \& {Bentley}}{{Zharkova} et~al.}{2003}]{2003SoPh..214...89Z}
{Zharkova} V.~V.,  {Ipson} S.~S.,  {Zharkov} S.~I.,  {Benkhalil} A.,
  {Aboudarham} J.,   {Bentley} R.~D.,  2003, \mn@doi [\solphys]
  {10.1023/A:1024081931946}, \href
  {https://ui.adsabs.harvard.edu/abs/2003SoPh..214...89Z} {214, 89}

\makeatother
\end{thebibliography}





\bsp	
\label{lastpage}
\end{document}